\documentclass[pra,twocolumn,superscriptaddress,showpacs]{revtex4}
\usepackage{amsmath}
\usepackage{amsfonts}
\usepackage{graphicx}   
  
\begin{document}
\title{Classical Dynamics as Constrained Quantum Dynamics}
\author{Stephen D. Bartlett}
\affiliation{Department of Physics, University of Toronto, 
  Toronto, Ontario M5S 1A7, Canada}
\affiliation{Department of Physics, Macquarie University, 
  Sydney, New South Wales 2109, Australia}
\author{David J. Rowe}
\affiliation{Department of Physics, University of Toronto, 
  Toronto, Ontario M5S 1A7, Canada}
\date{August 28, 2002} 

\begin{abstract} 
  We show that the classical mechanics of an \emph{algebraic model}
  are implied by its quantizations.  An algebraic model is defined,
  and the corresponding classical and quantum realizations are given
  in terms of a spectrum generating algebra.  Classical equations of
  motion are then obtained by constraining the quantal dynamics of an
  algebraic model to an appropriate coherent state manifold.  For the
  cases where the coherent state manifold is not symplectic, it is
  shown that there exist natural projections onto classical phase
  spaces.  These results are illustrated with the extended example of
  an asymmetric top.
\end{abstract}
\maketitle

\section{Introduction} 

The observables of a finite dimensional Lie algebra of an algebraic
system (defined below) are quantized by construction of an irreducible
unitary representation of that Lie algebra, in accordance with Dirac's
prescription~\cite{diracbook}.  Moreover, as discussed in two previous
papers~\cite{scalar,vector}, this quantization is achieved in a simple
and highly practical way by coherent state and vector coherent state
methods which, together with the theories of induced
representations~\cite{mackey} and geometric
quantization~\cite{kostant,souriau,woodhouse}, provide simple and
powerful techniques for quantizing complex systems.  However, it is
also known that the quantization of observables that do not belong to
a subalgebra of the infinite-dimensional Lie algebra of all
observables is impossible by Dirac's prescription~\cite{joseph}.  Even
for observables belonging to the universal enveloping algebra of a
finite-dimensional algebra of observables, there is the so-called
``ordering ambiguity''.  The method of geometric quantization provides
an elegant prescription for quantizing certain observables (those that
preserve a polarization), but does not provide quantizations for all
observables.  In this paper, we start from the premise that the
fundamental dynamics of physical systems are given by quantum
mechanics and proceed to show that the classical mechanics of an
algebraic system are implied by its quantizations.  This result shows
how classical mechanics can be defined within quantum mechanics and
establishes rules for the inverse process of quantization.  Thus, we
suggest that a criterion for a valid quantization is that it should be
consistent with dequantization.  A related but distinct problem is the
one of explaining why most macroscopic systems are observed to behave
classically; however, we do not address this problem.

Our starting point is the observation that a quantal Hilbert space is
a symplectic manifold and quantum mechanics is Hamiltonian mechanics
on such a manifold.  It is also known that restricting Dirac's
time-dependent variational principle,
\begin{equation}
  \delta \int \langle \psi(t) | \Big( \hat{H}-i\hbar
  \frac{\partial}{\partial t}\Big) |\psi(t)\rangle \, {\rm d}t = 0 \,, 
  \label{eq:Dirac}
\end{equation}
to a symplectic submanifold of the Hilbert space gives rise to
Hamiltonian equations of motion on that submanifold.  Approximate
Hartree--Fock theories~(cf.\ references cited in \cite{rowebook}),
theories of large amplitude collective motion~\cite{TDHF,rowe80} and
the density dynamics of Rowe, Vassanji and Rosensteel~\cite{rowe83}
have utilized this property of quantum systems extensively.  Indeed
many approximate many--body theories are known (cf.~\cite{rowe80}) to
be expressible in terms of Hamiltonian equations in a classical form
on symplectic submanifolds of the many--body Hilbert space.
Hartree--Fock theory, time--dependent Hartree--Fock theory, the
random--phase approximation, and the double--commutator
equations--of--motion method~\cite{rowebook} are all describable in
this context.  Thus, given a Hilbert space for a quantum system,
corresponding Hamiltonian equations of motion are defined by an
embedding of a symplectic phase space in this Hilbert space as a
submanifold and constraining the quantal dynamics to this submanifold.
In this paper, we show that quantum mechanics constrained to an
appropriate submanifold can lead to classical equations of motion, and
evolution of observables consistent with the corresponding classical
model.

For models with an algebraic structure, there is a straightforward and
transparent means to define corresponding quantal and classical
realizations using a spectrum generating algebra (SGA)~\cite{scalar}.
It is shown in this paper that classical phase spaces for an algebraic
model are embedded in its quantal representations.  This embedding
leads to a kinematical relationship between the symplectic structures
of the classical and quantum models.  The embeddings are often given
by coherent state submanifolds of the quantal Hilbert space in a
generalization of the well-known coherent states of the harmonic
oscillator.  In addition to the kinematical relationships given by
such embeddings, we show that constraining quantum dynamics to an
appropriate coherent state submanifold leads to classical equations of
motion for all the observables of the model; i.e., classical dynamics
is obtained as constrained quantum dynamics.  For coherent state
manifolds that are not symplectic, we consider their natural
projections onto classical phase spaces.  Thus, we show that it is
possible to regain the classical dynamics of a model from its
quantizations.  This result goes beyond Ehrenfest's
theorem~\cite{Ehr27,Bal94}  to give classical
equations of motion in terms of a classical Hamiltonian for \emph{all}
observables.

A potential ambiguity that arises is that there are many possible
embeddings of a classical phase space in a given Hilbert space.  It is
then important to enquire if different but equally reasonable
embeddings might give different classical dynamics.  We show that the
embedding problem is related to determining which quantal state of a
system is most appropriately assigned to a classically observed state.
Rather than a pure state, it seems natural to assign some mixture of
states, such as a thermal distribution with expectations of the
observables having values defined with distributions commensurate with
those observed.  However, with incomplete knowledge of a system, it is
clear that specification of its quantal state cannot be unique.  Thus,
if the classical dynamics that emerge from different choices were to
depend sensitively on the choice, it would be ambiguous.  It is
suggested in this paper by analyses of model systems that, while the
various classical dynamics given by constrained quantum mechanics are
not unique, they are nevertheless consistent with a single
\emph{ideal} classical mechanics defined as follows.

In an ideal classical mechanics, a point of the phase space is
identified with a state of a system whose observables (e.g., position
and momentum coordinates) all have precisely defined values.  If
quantum mechanics is fundamental, this description must be an
idealization because it does not accurately describe any physical
system obeying the uncertainty principle.  Thus, a realistic classical
description of a system should represent a state by a probability
distribution of ideal classical
states~\cite{Bal94}, with mean values and
variances that reflect these (classical) uncertainties.  Each state in
the probability distribution of ideal classical states would then
evolve in accordance with the idealized classical mechanics so that,
in any given situation, there would be a distribution of possible
outcomes.  We shall refer to such a dynamics as \emph{the physical}
classical dynamics.

It is of fundamental interest for the interpretation of quantum
mechanics to be aware that many of its superficial differences with
classical mechanics result from comparing it with the ideal rather
than a physical version of the classical theory.  The familiar example
of a playing card balanced on one end on a flat table illustrates this
point~\cite{Teg01}.  According to ideal classical mechanics, the card
is in an equilibrium configuration and in the absence of interactions
with its environment (e.g., air currents), it should remain in this
configuration indefinitely.  However, in quantum mechanics, its wave
function is not in a stationary state and it will evolve symmetrically
in such a way that the card is predicted to fall, with equal
probability, to one side or the other.  Exactly the same conclusion is
reached in a physical classical analysis in which the initial state of
the card is described by a symmetrical distribution of configurations
about equilibrium.  We do not suggest that quantum and classical
mechanics necessarily give similar results, but simply stress that it
is only meaningful to compare a physical classical mechanics with a
fundamentally quantum description.

This paper is structured as follows.  Sec.~\ref{sec:Barrier} presents
the familiar example of barrier penetration to illustrate the concepts
of ideal and physical classical dynamics, and how they relate to
quantum and constrained quantum dynamics.  In
Sec.~\ref{sec:background}, we introduce a description of both quantum
and classical algebraic models in terms of an SGA, and formulate the
concept of densities to describe a state of the classical system.  In
Sec.~\ref{sec:Quantum}, we investigate how constraining the dynamics
to appropriate coherent state manifolds can lead to classical
equations of motion.  These techniques are illustrated in
Sec.~\ref{sec:Rotor} with a nontrivial application to the dynamics of
an asymmetric top.  Conclusions are presented in
Sec.~\ref{sec:Conclusions}.

\section{Example:  Barrier penetration}
\label{sec:Barrier}

Barrier penetration is often used to illustrate the differences
between classical and quantum mechanics.  In this section, we use a
barrier penetration example to introduce the
essential concepts and principles that will be developed in the
remainder of the paper.

We consider a point particle in one dimension, with Hamiltonian
\begin{equation}
  \label{eq:CanonicalHam}
  {H} = \frac{{p}^2}{2m} + {V} \, ,
\end{equation}
containing a potential energy defined by
\begin{equation}
  \label{eq:PotentialBarrier}
  V(x) = \begin{cases} V_0 \quad& 0 \leq x \leq L \, , \\
  0&{\rm otherwise} \, , \end{cases}
\end{equation}
where $V_0$ and $L$ are positive real numbers.  For simplicity, we
work in units in which $\hbar^2/2m =1$ and, in these units,
set the barrier height to $V_0=1$.  The
momentum of the particle is then $p=\hbar k$, where $k$ is the wave
number, and the energy of the particle, when outside of the barrier,
is its kinetic energy $E(k)=k^2$.

\subsection{Ideal versus physical states in classical mechanics}

In ideal classical mechanics, the state of a particle in
this system  is represented as a point of position $x$
and momentum $p$ in a classical $(x,p)$ phase space.

In a description that more accurately represents a physical classical
system, a state of the particle is represented by a probabililty
distribution $P^{\rm CM}(\bar x,\bar k; x,k)$ of ideal classical
states having mean position and momentum, given by
\begin{align}
  \bar x &= \langle x\rangle = \int x P^{\rm CM}(\bar x,\bar k; x,k)
  \,{\rm d}x\, {\rm d}k \,,\nonumber\\
  \bar p &= \langle \hbar k\rangle = \int \hbar k P^{\rm CM}(\bar x,\bar k;
   x,k) \,{\rm d}x\, {\rm d}k \,,
\end{align}
and corresponding variances in these means
\begin{align}
  \sigma^2(\bar x) &=  \int (x-\bar x)^2 P^{\rm CM}(\bar x,\bar k; x,k)
  \,{\rm d}x\, {\rm d}k \,,\nonumber\\
  \sigma^2(\bar p) &=  \int (\hbar k - \bar p)^2 P^{\rm CM}(\bar x,\bar
  k; x,k) \,{\rm d}x\, {\rm d}k \,.
\end{align}

To be specific, we consider classical probability distribution functions 
given by 
\begin{equation}
  P^{\rm CM}_\alpha(\bar x, \bar k; x,k) = \frac{1}{\pi} e^{-(x-\bar
  x)^2 /\alpha} e^{- \alpha(k-\bar k)^2}\,.
  \label{eq:gaussian2}
\end{equation}
for  various values of $\alpha$.  With
physical states characterized by such distributions, the measured
values of observables, such as the potential energy $V$, would have
expectation values given, for example, by
\begin{equation}
  \bar V_\alpha(\bar x) = \int V(x) \, 
  P^{\rm CM}_{\alpha} (\bar x,\bar k;
  x,k)\,{\rm d}x\,{\rm d}k \,, 
  \label{eq:Vexpectation}
\end{equation}
and there would be corresponding uncertainties in these measured
values due both to the errors of the measuring apparatus as well as
the fundamental uncertainties in the position of the particle in a
distribution of ideal states. As an illustration, Fig.~1 shows that
expectation values of the potential barrier for the classical
probability distribution functions given by Eq.~(\ref{eq:gaussian2})
for various values of $\alpha$.

\begin{figure}
  \includegraphics*[height=3in,keepaspectratio]{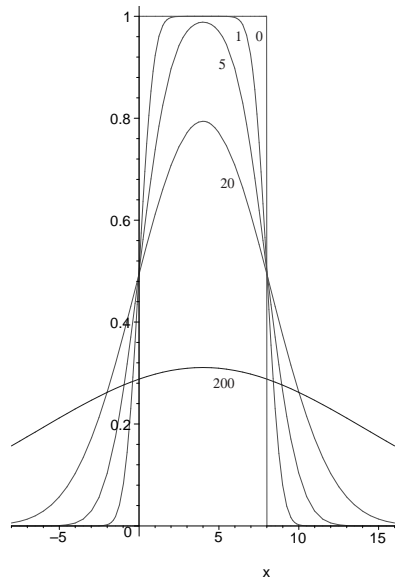}
  \caption{Expectations values $\bar V_\alpha$, defined by
    Eq.~(\ref{eq:Vexpectation}), of the potential,
    Eq.~(\ref{eq:PotentialBarrier}), for physical classical states
    having Gaussian probability distributions in position and momentum
    defined by Eq.~(\ref{eq:gaussian2}) with various values of
    $\alpha$.  The same expectation values are obtained for minimal
    uncertaintly quantum mechanical states with wave functions given
    by Eq.~(\ref{eq:minwfn}). \label{fig:1}} 
\end{figure}     

It should be noted that the classical probability distribution $P^{\rm
  CM}_\alpha(\bar x,\bar k;x,k)$ represents a physical state having
the minimal product of uncertainties in position and momentum allowed
by quantum mechanics.  In general, physical states have much greater
products of uncertainties.

The accuracy in the measurement of an observable function of position can
be increased without limit, in principle, by admitting a
correspondingly larger uncertainty in momentum.  Thus, by letting
$\alpha\to 0$, there is no limit in principle on the accuracy with
which the potential energy or some other single observable can be
measured. However, one cannot simultaneously measure the values of
different (non-commuting) observables that depend on both the position
and the momentum of the particle to greater precision than allowed by
the uncertainty principle  in any system with a
fundamentally quantum description. 

\subsection{Constrained and unconstrained states in quantum mechanics}

A state $|\psi\rangle$ of a particle in quantum mechanics defines
probability distributions in the position and momentum variables
given, respectively, by the square moduli of its wave functions
$|\psi(x)|^2  = |\langle x|\psi\rangle|^2$ and
$|\varphi(k)|^2  = |\langle k|\psi\rangle|^2$
in position and momentum representations.  In particular, the mean
values and variances of the position and momentum variables in a state
$|\psi\rangle$ are given by
\begin{equation}
  \bar x = \langle \psi|\hat x|\psi\rangle \,, \quad
  \sigma^2(\bar x) = \langle \psi|(\hat x - \bar x)^2|\psi\rangle \,, 
\end{equation} 
and
\begin{equation}
  \bar p = \langle \psi|\hat p|\psi\rangle\,, \quad
  \sigma^2(\bar p) = \langle \psi|(\hat p-\bar p)^2|\psi\rangle \,, 
\end{equation}
where $\hat x$ and $\hat p$ are the quantal position and momentum
operators; in the position representation,
they are given by $\hat x=x$ and $\hat p = -i\hbar \partial/\partial
x$.

In constrained quantum mechanics, the dynamics is restricted to a
submanifold of states distinguished by their mean values of $\bar x$
and $\bar p =\hbar \bar k$. A particular submanifold can be selected
in many ways. For illustrative purposes, we consider here a set of
minimum uncertainty states $\{ |\alpha,\bar x, \bar k\rangle\}$ (of
fixed $\alpha$) with wave functions $\{\psi(\alpha,\bar x,\bar k)\}$
 in the position representation 
\begin{equation}
  \psi(\alpha,\bar x,\bar k; x) =  (\pi\alpha)^{-1/4} e^{-(x-\bar
  x)^2/2\alpha} e^{i\bar k x} \,. 
  \label{eq:minwfn}
\end{equation}
These wave functions have probability distribution in $x$ given by
\begin{equation}
  P^{\rm QM}_\alpha(\bar x;x) = \frac{1}{\sqrt{\pi\alpha}} e^{-(x-\bar
  x)^2/\alpha} \,.
\end{equation}
The corresponding wave functions in the momentum  representation are
given by the Fourier transforms
\begin{equation}
  \varphi(\alpha,{\bar x},\bar k; k) = (\alpha/\pi)^{1/4} e^{-\alpha
  (k-\bar  k)^2/2} e^{-i\bar xk}\,.
\end{equation}
Hence, the probability distributions in momentum $p=\hbar k$ are given
for these states by
\begin{equation}
  P^{\rm QM}_\alpha(\bar k;k) = \sqrt{\frac{\alpha}{\pi}}\,
  e^{-\alpha (k-\bar  k)^2} \,.
\end{equation}
It is seen that the products of these distributions are identical to
the classical distributions of Eq.~(\ref{eq:gaussian2}), i.e.,
\begin{equation}
  P^{\rm CM}_\alpha(\bar x, \bar k;x,k) =  P^{\rm QM}_\alpha(\bar x;x)
  P^{\rm QM}_\alpha(\bar k;k) \,.
\end{equation}
Thus, the expectation values of, for example, the potential energy in
a state $|\alpha,\bar x,\bar k\rangle$ are given by
Eq.~(\ref{eq:Vexpectation}) with $ P^{\rm CM}(\bar x, \bar k;x,k) =
P^{\rm QM}_\alpha(\bar x;x) P^{\rm QM}_\alpha(\bar k;k)$.

This example illustrates the fact that a suitably selected set of
constrained quantum mechanical states can give precisely the same mean
values and variances as a corresponding set of physical classical
states.  It also makes clear that in suitable situations, e.g.\ when
the Hamiltonian is a sum of a potential energy that is a function of
only position coordinates and a kinetic energy that is a function of
only momenta, it is possible (at least in principle) to probe the
functional forms of the components of the Hamiltonian separately in
both physical classical mechanics and contrained quantum mechanics to
any desired accuracy.  The important conclusion is that by
observations of the values of physical observables in quantum
mechanics it is possible to make precisely the same inferences about
the ideal classical expressions of observables, e.g.\ as functions of
position and momentum, as is, in principle, possible in physical
classical mechanics. Of course, it is fully recognized that such a
claim is not established by consideration of a single example.
However, we suggest that the validity of
this claim should follow from a suitably precise definition of what is
meant by physical classical states.

\subsection{Barrier penetration in classical mechanics} 

We now illustrate the meaning of \emph{ideal} and \emph{physical}
classical mechanics in the context of the barrier penetration example
and then do the same for \emph{unconstrained} and \emph{constrained}
quantum mechanics.

In ideal classical mechanics, the probability for penetration of the
barrier by a particle approaching the barrier with a precisely defined
momentum $\hbar k$ is simply
\begin{equation}
  T_{\rm ICM}(k) = \begin{cases}1& {\rm if}\; k>1,\\ 0 & {\rm if}\;
  k<1.\end{cases}
\end{equation}
Thus, in a physical situation in which the incident particle state at
time $t=0$ is in a distribution of ideal classical states given by a
function $P^{\rm CM}_\alpha(\bar x, \bar k;x,k)$, the probability for
penetration of the barrier is given by
\begin{equation}
  T_{\rm CM}(\alpha,\bar k) = \int P^{\rm CM}_\alpha(\bar x,\bar
  k;x,k)\,  T_{\rm ICM}(k)\,{\rm d}x\, {\rm d}k \,.
\end{equation} 

\begin{figure*}
  \includegraphics*[height=3in,keepaspectratio]{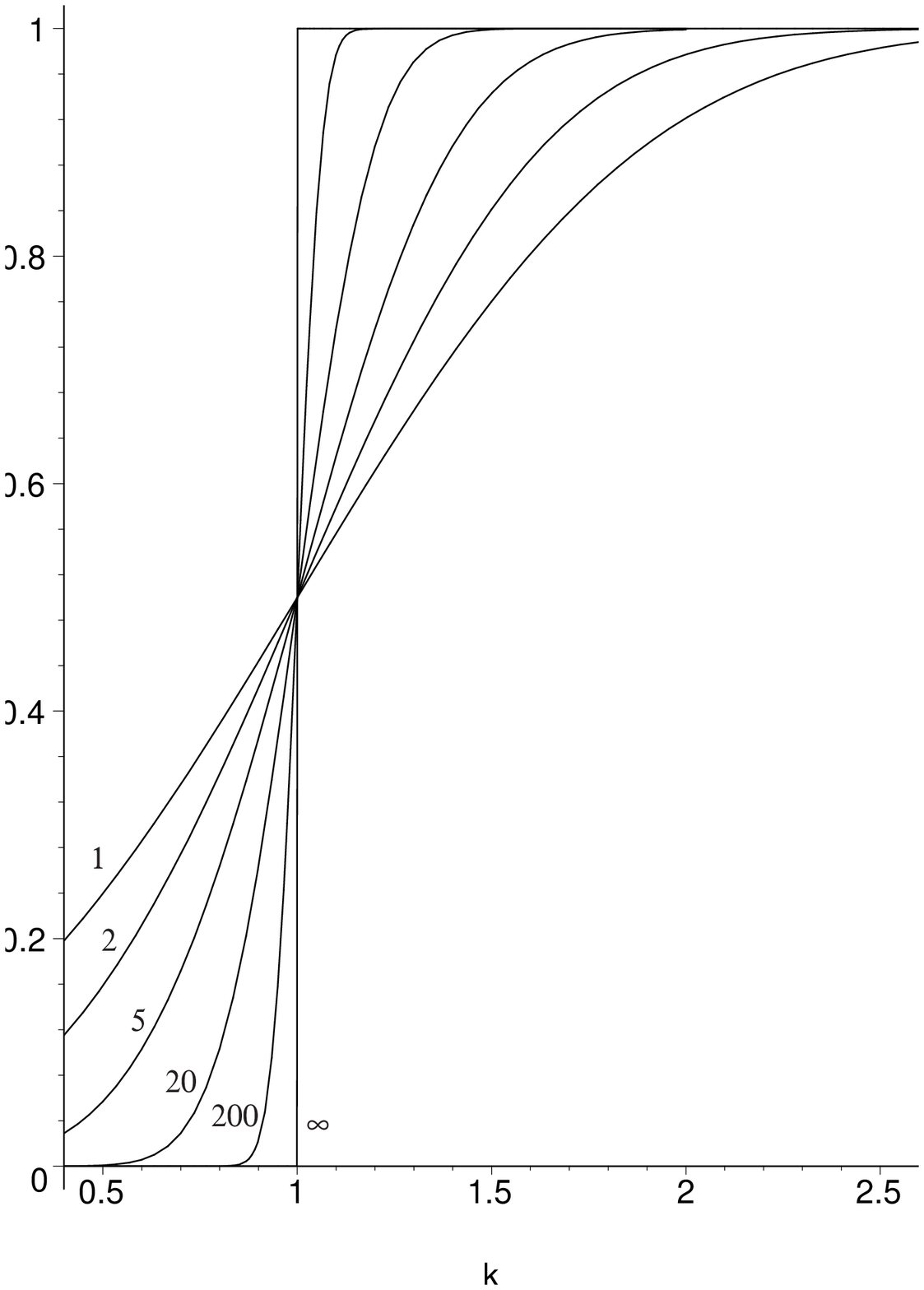}
  \hspace{2cm}
  \includegraphics*[height=3in,keepaspectratio]{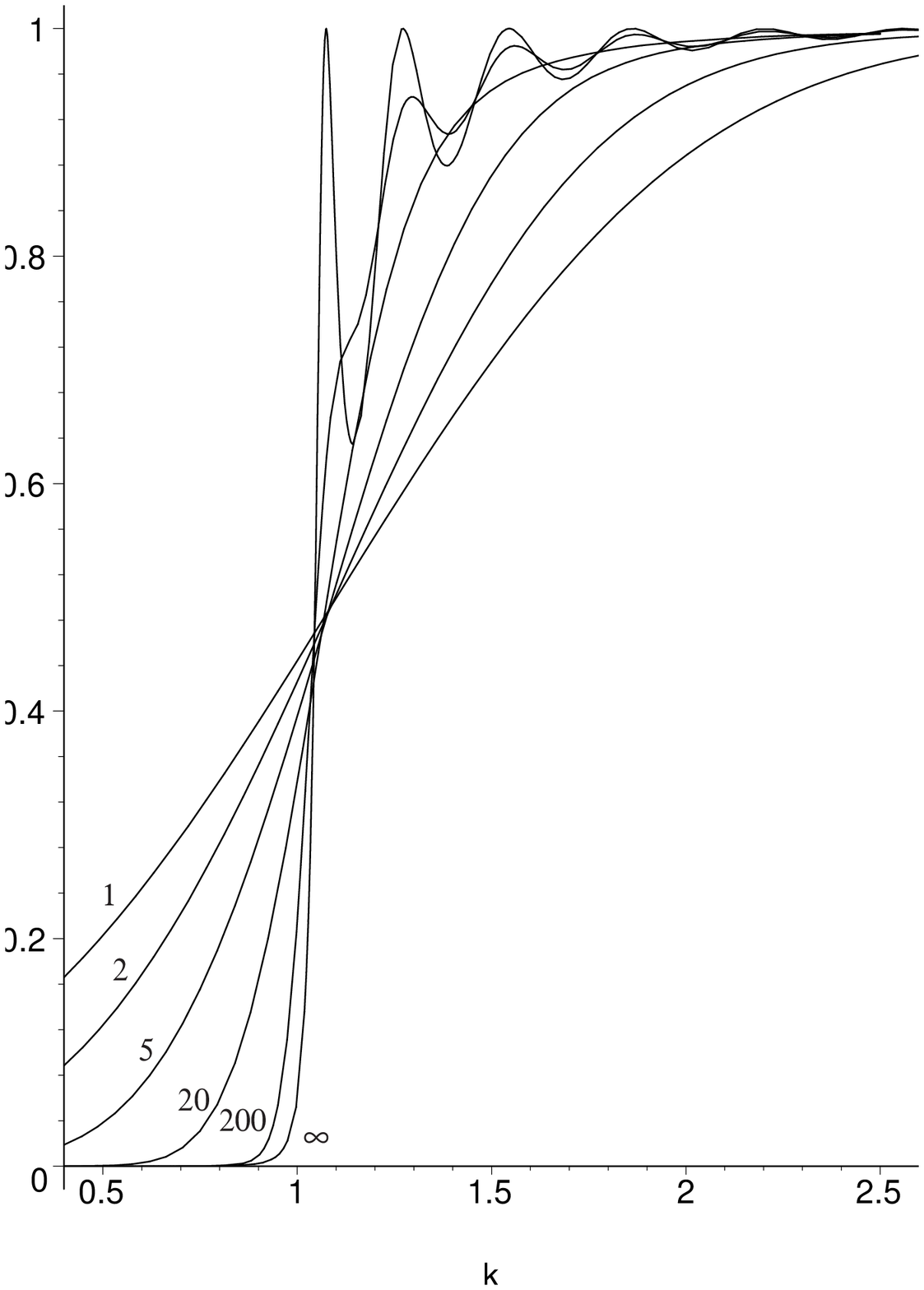}
  \caption{Transmission probabilities in physical classical
    mechanics (left) for Gaussian probability distributions in
    position and momentum defined, by Eqs.~(\ref{eq:gaussian2}) and
    for corresponding minimal wave packets in quantum mechanics
    (right). The results are shown for a range of values of the
    Gaussian width parameter $\alpha$. \label{fig:2}}
\end{figure*}

Numerically integrated values for these barrier penetration
probabilities are shown as functions of the wave number $k$, for
different values of $\alpha$, in Fig.~\ref{fig:2} for the particular
case of a barrier of width $L=8$.

It is seen that, in the limit as $\alpha\to\infty$ and the momentum of
the particle becomes precisely defined, the barrier penetration
probability approaches that of ideal classical mechanics; this
reflects the limiting value
\begin{equation}
  \lim_{\alpha\rightarrow\infty} T_{\rm CM}(\alpha,k) = T_{\rm ICM}(k)
  \,. 
\end{equation}

\subsection{Barrier penetration in quantum mechanics}

In quantum mechanics, the probability for penetration of the barrier
in the limit in which the incident particle is in a momentum
eigenstate is given by
\begin{equation}
  T_{\rm QM}(k) = \begin{cases}\displaystyle
  \frac{8k^2(k^2-1)}{8k^2(k^2-1) +1- \cos(2L(k^2-1)) }{\rm if}\; k>1,\\
  \displaystyle
  \frac{8k^2(k^2-1)}{8k^2(k^2-1) +1- \cosh(2L(1-k^2)) }{\rm if}\; k<1.
  \end{cases}
\end{equation}
This function is shown as the $\alpha = \infty$ curve on the right of
Fig.~\ref{fig:2} for a barrier of width $L=8$.

For a finite value of $\alpha$, the penetration probability for a
particle in a state $|\alpha,\bar x, \bar k\rangle$ at time $t=0$ is
given by
\begin{equation}
 T_{\rm QM}(\alpha,\bar k) = \int T_{\rm QM}(k) P^{\rm QM}_\alpha(\bar
  k;k)\,{\rm d}k \,.
\end{equation}
The barrier penetration probabilities $T_{\rm QM}(\alpha,
\bar{k} )$ are shown
for a range of value of $\alpha$ in Fig.~\ref{fig:2}.  In parallel
with the classical probabilities, the quantal penetration
probabilities approach those for which the momentum is precisely
defined in the $\alpha\to\infty$ limit.

It is notable that the quantal penetration probabilities have a
remarkable resemblance to their classical counterparts.  On
examination, it is found that the classical penetration of the barrier
exceeds that of quantum mechanics for all but a small region of
$k\lesssim 1$. The remarkable feature of the
quantal treatment is not so much the penetration of the barrier when
the energy is less than would be required classically, as the
reflection that occurs when the half wave length is an integer
fraction of the barrier width.

\subsection{Constrained quantum mechanics}

Constrained quantum mechanics, as we define it in
Sec.~\ref{sec:Quantum}, gives the time evolution of a
state subject to the constraint that it remains a coherent state at
all times. As will be shown generally in this paper, the time
evolution of the constrained wave function in the present example is
 of the form
\begin{equation}
  \psi(x,t) = \psi(\alpha, {\bar x}(t),\bar k(t);x)\,e^{iE_\alpha t} \,,
\end{equation}
where $\psi(\alpha,{\bar x},\bar k ;x)$ is given by
Eq.~(\ref{eq:minwfn}) and $E_\alpha$ is the energy expectation of the
coherent state. Thus, the time evolution of $\psi(x,t)$ is defined by
the time evolutions of $\bar x(t)$ and $\bar k(t)$. We show, that
$\bar x(t)$ and $\bar p(t)=\hbar \bar k(t)$ are given by classical
dynamics for the Hamiltonian
\begin{equation}
  H_\alpha ( x,k) = \langle \psi(\alpha, x,
  k)|\hat H |\psi(\alpha,x,k)\rangle \,,
  \label{eq:HfromQM}
\end{equation}
where $\hat H$ is the quantal Hamiltonian given by replacing the
momentum $p$ in the classical expression (\ref{eq:CanonicalHam}) by
the usual quantization $\hat p = -i\hbar {\rm d}/{\rm d}x$.  One
obtains
\begin{equation}
  H_\alpha (x,k) = k^2 +
  \bar V_\alpha( x) +{\rm const.}\,,
\end{equation}
where the potential $\bar V_\alpha ( x)$ is defined by
Eq.~(\ref{eq:Vexpectation}). Thus, the probability for penetration of
the barrier given by constrained quantum mechanics becomes identical
to that of ideal classical mechanics provided the maximum value of the
potential $ \bar{V}(x)$ is the same as that for the original potential
$V(x)$.  The maxima are the same to a high degree of accuracy (cf.\ 
Fig.~\ref{fig:1}) for $\alpha \lesssim 2$ and become identical in the
limit as $\alpha\to 0$, i.e., the limit in which the state
$|\psi(\alpha,{\bar x},\bar k)\rangle$ becomes an eigenstate of the
position operator $\hat x$.

In a general situation, constrained quantum mechanics is 
governed by  a Hamiltonian that is averaged, in parallel
with Eq.~(\ref{eq:HfromQM}), over the distributions of suitably
defined coherent states for the system under consideration. Thus, in
general, constrained quantum mechanics does not reproduce the original
Hamiltonian precisely (except for the harmonic oscillator and limiting
cases).  But, by the same token, it is important to recognize that
neither is the Hamiltonian of ideal classical mechanics reproduced
exactly by physical classical mechanics except for similar limiting
cases.  Indeed, constrained quantum mechanics gives exactly the same
Hamiltonian as physical classical mechanics when averaged over the
same position and momentum distributions.  That one does not reproduce
the ideal classical Hamiltonian in either constrained quantum
mechanics or in physical classical mechanics is a direct reflection of
the limitation on observations imposed by the uncertainty principle.

Thus, while we cannot claim to derive the ideal classical mechanics of
a system from its quantizations, we can claim to derive a classical
mechanics that is consistent with the ideal classical mechanics in a
sense that is similar to the way that physical classical mechanics is
consistent with, but not identical to, ideal classical mechanics.

We consider these limitations in deriving the ideal classical
dynamics of a system from quantum mechanics to be fundamental and to
reflect the physics of the uncertainty principle.  In particular, we
claim that any inferences about relationships between observables that
can be obtained by physical measurement can also be inferred as
precisely as the uncertainty principle allows by quantum mechanical
considerations.

\section{Classical and quantum algebraic models}
\label{sec:background}

In this section, we review the background material for describing an
algebraic model, both classically and quantally, with a focus on the
kinematical structure.  For further details on algebraic models,
see~\cite{scalar}.  Also, Marsden and Ratiu~\cite{marsden94} provide
details on coadjoint orbits, Hamiltonian actions, and Hamiltonian
formulations of quantum mechanics.

\subsection{Observables and spectrum generating algebras} 
\label{sec:SGA}

In classical mechanics, observables are realized as smooth
real--valued functions on a connected phase space $\mathcal{M}$, i.e.,
elements of $C^{\infty}(\mathcal{M})$.  They form an
infinite--dimensional Lie algebra with Lie product given by a Poisson
bracket.  In quantum mechanics, observables are interpreted as
Hermitian linear operators on a Hilbert space $\mathbb{H}$; they are
elements of $GL(\mathbb{H})$ and form an infinite--dimensional Lie
algebra with Lie product given by commutation. 

The algebras $C^{\infty}(\mathcal{M})$ and $GL(\mathbb{H})$ for a
given physical system are different~\cite{joseph}.  However, for an
algebraic system (defined below) it is possible to establish a simple
relationship between finite--dimensional subalgebras of
$C^{\infty}(\mathcal{M})$ and $GL(\mathbb{H})$.  Let $\mathfrak{g}$
denote an abstract Lie algebra of observables that is real and
finite--dimensional. Suppose that $\mathfrak{g}$ can be represented
classically by a homomorphism $J:\mathfrak{g}\to
C^{\infty}(\mathcal{M})$ and quantum mechanically by a unitary
representation $T:\mathfrak{g} \to GL(\mathbb{H})$.  Let $\mathcal{A}
= J(A)$ and $\hat A = T(A)$ denote the classical and quantal
representations, respectively, of an element $A\in\mathfrak{g}$. Then,
if elements $A$, $B$, and $C\in \mathfrak{g}$ satisfy the commutation
relations
\begin{equation}
  \label{eq:AbstractAlgebraLieBracket}
  [A,B] = i \hbar \, C \, ,
\end{equation}
the corresponding linear operators and functions satisfy
\begin{equation}
  [\hat A,\hat B] = i \hbar \, \hat C \, ,
\end{equation}
and
\begin{equation}
  \{ \mathcal{A} , \mathcal{B} \} = \mathcal{C} \, ,
\end{equation}
where $\{ \; ,\;\}$ denotes the classical Poisson bracket.  (More
precisely, the homomorphism is given by $\{ (i\hbar\mathcal{A}) ,
(i\hbar\mathcal{B}) \} = i\hbar (i\hbar\mathcal{C})$.)

It should be emphasized that the presence of the $i\hbar$ factor in
the commutation relation, Eq.~(\ref{eq:AbstractAlgebraLieBracket}), of
the abstract algebra has no quantum mechanical implications.  The
factor $\hbar$, for example, {can be} regarded simply as a suitable
unit, chosen such that the Lie bracket $[A,B]$ has the same dimensions
(i.e., is expressed in the same units) as a simple product of $A$ and
$B$.  The Poisson bracket, which differentiates, e.g., with respect to
$x$ and $p$, does not have this property.  The factor $i\hbar$ of the
Lie bracket can be removed by simply dividing each of $A$, $B$ and $C$
by $i\hbar$.  The dependence on $\hbar$ (equivalent to setting $\hbar
=1$) can also be removed by expressing the observables in any
convenient dimensionless units.

Let $\mathcal{G}\subset C^{\infty}(\mathcal{M})$ denote the classical
algebra $\mathcal{G} =\{ J(A) | A\in \mathfrak{g}\}$.  If the values
of the observables in $\mathcal{G}$ are sufficient to uniquely
identify a point in $\mathcal{M}$, the algebra $\mathcal{G}$ (or
$\mathfrak{g}$) is said to be a \emph{spectrum generating algebra\/}
(SGA) for the classical system~\cite{ihrig}.  The Lie algebra
$\mathfrak{g}$ is said to be a SGA for a quantal system if the Hilbert
space for the system carries a unitary irreducible representation of
$\mathfrak{g}$~\cite{SGA}.  A model dynamical system having a
finite--dimensional SGA is said to be an \emph{algebraic
  system}~\cite{scalar}.

Note that the SGA does not necessarily include the Hamiltonian.  In
fact, {in a generic situation,} the Hamiltonian is \emph{not} an
element of the SGA.  To be useful, one may require that the
Hamiltonian and other important observables of the system should be
simply expressible in terms of $\mathfrak{g}$, e.g., by belonging to
its universal enveloping algebra. 

It is tempting to infer from the above considerations that the desired
maps between the classical and quantal realizations of an algebraic
system are simply Lie algebra homomorphisms.  This idea underlies
Dirac's canonical quantization~\cite{diracbook}.  However, these
realizations do not give a complete description of the relationship
between a quantum and classical model.  In the first place, there may
be many classical and many quantal representations of a given SGA but
only some quantal representations qualify as quantizations of a given
classical representation, and vice versa.  Moreover, as proven by the
famous Groenwald--van Hove theorem (see~\cite{gotay}), the full
algebra of classical observables has no irreducible representations.

\subsection{Classical phase spaces as coadjoint orbits}

The phase space of a classical algebraic model is naturally viewed as
a coadjoint orbit of a dynamical group.  Coadjoint orbits are
mathematical constructions that have been widely studied and are known
to have many useful properties~\cite{marsden94}.  In particular, they
are known to be symplectic manifolds.  For further details of the
following construction, see~\cite{scalar}.

Let $G$ be a group of canonical transformations of a phase space
$\mathcal{M}$ for a classical model.  Then, if $G$ acts transitively
on $\mathcal{M}$, it is said to be a \emph{dynamical group} for the
model.  If an element $g\in G$ sends a point $m\in\mathcal{M}$ to
$m\cdot g\in\mathcal{M}$, then $\mathcal{M}$ is the group orbit
\begin{equation}
  \mathcal{M} = \{ m\cdot g\, |\, g\in G\} 
\end{equation}
and diffeomorphic to the factor space $H_m\backslash G$
with isotropy subgroup
\begin{equation}
  \label{eq:PhaseSpaceIsotropySubgroup}
  H_m = \{ h\in g \,|\, m\cdot h = m\} \, .
\end{equation}
A remarkable fact~\cite{marsden94} that will be used extensively in
the following is that a phase space with a dynamical group $G$ can be
identified with a coadjoint orbit of $G$.  Conversely, every coadjoint
orbit is a phase space.  Moreover, the Lie algebra $\mathfrak{g}$ of
$G$ is a SGA for the model.

Recall that $G$ has a natural adjoint action on its Lie algebra
$\mathfrak{g}$
\begin{equation}
  {\rm Ad}(g) : \mathfrak{g} \to \mathfrak{g} ;\ A \mapsto A(g)=
  {\rm Ad}(g) A \, , \label{eq:Adg}
\end{equation}
where, for a matrix group, ${\rm Ad}(g) A = gAg^{-1}$.  $G$ also has a
coadjoint action on the space $\mathfrak{g}^*$ of real--valued linear
functionals on $\mathfrak{g}$ (the dual of $\mathfrak{g}$).  Thus, if
$\rho$ is an element of $\mathfrak{g}^*$ and $\rho_g$ is defined by
\begin{equation}
   \rho_g(A) = \rho(A(g)) \,,\quad \forall\, A\in \mathfrak{g}\,,
\end{equation}
then the coadjoint orbit
\begin{equation}
  \mathcal{O}_\rho = \{ \rho_{g} \, |\, g\in G \} \, ,
\end{equation}
is diffeomorphic to the factor space $H_\rho\backslash G$ with
isotropy subgroup $H_\rho = \{ h\in g \,|\, \rho_h = \rho\}$.  We
shall refer to an element $\rho$ of $\mathfrak{g}^*$ as a
\emph{classical density}.  Now if a density
$\rho\in\mathfrak{g}^*$ is chosen such that $H_\rho=H_m$ then there is
a diffeomorphism $\mathcal{M}\to \mathcal{O}_\rho$ in which $m \mapsto
\rho$ and $m\cdot g \mapsto \rho_g$.  This map is known as a
\emph{moment map}.

The map $\mathcal{M}\to \mathcal{O}_\rho$ defines a classical
representation $J:\mathfrak{g}\to\mathcal{G}; A\mapsto \mathcal{A}=
J(A)$ of the Lie algebra $\mathfrak{g}$ as functions over the
classical phase space $\mathcal{M}$, defined by
\begin{equation}
 \mathcal{A}(m\cdot g) = \rho_g (A)\,,
\end{equation}
with Poisson bracket given by
\begin{equation}
  \label{eq:PoissonBracket}
  \{\mathcal{A},\mathcal{B}\} = \sigma (A,B) \, ,
\end{equation}
where $\sigma$ is the antisymmetric two--form on the algebra with
values at $m\cdot g \in \mathcal{M}$ given by
\begin{equation}
  \label{eq:OmegaAsDRho}
  \sigma_{m\cdot g} (A,B) = - \frac{i}{\hbar} \, \rho_g ([A,B])
  \,,\quad \forall\, A,B\in \mathfrak{g}\,.
\end{equation}
This two--form is nondegenerate, and thus (with the realization of the
Lie algebra $\mathfrak{g}$ as a set of invariant vector fields on
$\mathcal{M}$) defines a symplectic form on $\mathcal{M}$.

The moment map $\mathcal{M} \to \mathcal{O}_\rho$ relates the
symplectic structure of these spaces, making them
\emph{symplectomorphic}.  Thus, $\sigma$ can be equivalently viewed as
a symplectic form on either $\mathcal{M}$ or $\mathcal{O}_m$; they are
equivalent as classical phase spaces.  This powerful
result allows us to interchange between the phase space $\mathcal{M}$
of an algebraic model and it's corresponding coadjoint orbit
$\mathcal{O}_m$; they are equivalent.

\subsection{Quantum mechanics as Hamiltonian mechanics}

A quantum system consists of a Hilbert space $\mathbb{H}$ and a set of
observables, including a Hamiltonian $\hat{H}$, given as Hermitian
linear operators in $GL(\mathbb{H})$.  Quantum dynamics is given by
the Schr\"{o}dinger equation
\begin{equation}
  \label{eq:SchrodingerEquation}
  i\hbar \frac{\partial}{\partial t} |\psi(t)\rangle = \hat{H}
  |\psi(t)\rangle \, ,
\end{equation}
for $|\psi(t)\rangle \in \mathbb{H}$.  In this section, quantum
dynamics is expressed as Hamiltonian dynamics relative to a natural
symplectic form on the corresponding projective Hilbert space
$\mathbb{PH}$.  (A projective Hilbert space is a Hilbert space
together with an equivalence relation which identifies vectors that
are the same to within a complex factor.  Thus, we say that
$|\psi\rangle \equiv |\phi\rangle$ if $|\psi\rangle = c |\phi\rangle$
for some $c\in\mathbb{C}$.)

The complex Hermitian inner product of a Hilbert space leads to a
Hermitian metric on the projective Hilbert space known as the
\emph{Fubini-Study metric}~\cite{marsden94}.  This metric provides two
real non-degenerate bilinear forms on $\mathbb{PH}$.  For $\mathbb{H}$
finite--dimensional, the first is defined in terms of a coordinate
patch $\{ \zeta^{\mu} \}$ by
\begin{equation} 
  g_{\mu\nu}(\psi) = g\big(\frac{\partial}{\partial
  \zeta^{\mu}},\frac{\partial}{\partial \zeta^{\nu}}\Big)(\psi) = {\rm
  Re} \Bigl\langle \frac{\partial \psi}{\partial \zeta^{\mu}} \Bigl|
  \frac{\partial \psi}{\partial \zeta^{\nu}} \Bigr\rangle \, .
\end{equation}
It is symmetric, hence Riemannian, and provides concepts of
distance and curvature. The second,
\begin{equation}
  \label{eq:DefFubiniStudy}
  \omega_{\mu\nu}(\psi) = \omega\big(\frac{\partial}{\partial
  \zeta^{\mu}},\frac{\partial}{\partial \zeta^{\nu}}\Big)(\psi)
  = -2 \hbar\ \text{Im}
  \Bigl\langle \frac{\partial \psi}{\partial \zeta^{\mu}} \Bigl|
  \frac{\partial \psi}{\partial \zeta^{\nu}} \Bigr\rangle \, ,
\end{equation}
is anti-symmetric and closed; hence it is symplectic. The latter
provides the basic structure whereby quantum mechanics can be
expressed in Hamiltonian form.  

If the Hilbert space $\mathbb{H}$ carries a unitary representation $T$
of a Lie group $G$, then the action of $G$ on $\mathbb{PH}$ leaves the
Fubini--Study metric invariant~\cite{marsden94}.  Thus, $G$ acts on
$\mathbb{PH}$ as a group of canonical transformations relative to the
symplectic form $\omega$. 

There exists a map from operators in $GL(\mathbb{H})$ to functions on
$\mathbb{PH}$ in which $\hat{F}$ maps to a function $F$ on
$\mathbb{PH}$ defined by its expectation value
\begin{equation}
   \label{eq:FunctionOnProjHilbert}
   F (\psi) = \langle{\psi}|{\hat{F}}|{\psi}\rangle \, .
\end{equation}
With the energy function $H$ defined on $\mathbb{PH}$ by the
expectation of the quantal Hamiltonian $\hat{H},$
\begin{equation}
  \label{eq:QuantumHamiltonianAsFunction}
  H(\psi) = \langle{\psi}|{\hat{H}}|{\psi}\rangle \, ,
\end{equation}
and its derivatives expressed
\begin{align}
  \frac{\partial H}{\partial \zeta^\nu} &=
  \left[  \Bigl\langle \frac{\partial\psi}{\partial
  \zeta^\nu}\Big|\hat H {\psi}\Big\rangle + \Bigl\langle{\hat H \psi}
  \Big| \frac{\partial\psi}{\partial \zeta^\nu}\Big\rangle \right]
  \nonumber \\
  &=i\hbar \left[  \Bigl\langle \frac{\partial\psi}{\partial
  \zeta^\nu}\Big|\dot{\psi}\Big\rangle - \Bigl\langle{\dot\psi}
  \Big|\frac{\partial\psi}{\partial \zeta^\nu}\Big\rangle \right] \, ,
\end{align}
with
\begin{equation}
  \dot\psi = \frac{\partial \psi}{\partial \zeta^\mu} \dot \zeta^\mu ,
\end{equation}
the time derivatives of the coordinates are determined~\cite{rowe80}
to be given by
\begin{equation}
  \dot{\zeta}^{\mu}  =
  \omega^{\mu\nu}\frac{\partial H}{\partial \zeta^\nu} \, .
\end{equation}
Here, $\omega^{\mu\nu}$ is the inverse of the symplectic metric
$\omega_{\mu\nu}$, defined by $\omega_{\mu\alpha}\omega^{\nu\alpha} =
\delta_\mu^\nu$.  Hence, the time evolution of $F$ is given by
\begin{equation}
  \label{eq:HamiltonianMechanicsOnHilbert}
  \dot{F} = \{ F , H \}_{\text{QM}} \, ,
\end{equation}
where the Poisson bracket of any two functions $F_1,
F_2$ on $\mathbb{PH}$ is defined by
\begin{equation}
  \label{eq:PoissonBracketOnProjHilbert}
  \{ F_1 , F_2 \}_{\text{QM}} =
  \frac{\partial F_1}{\partial \zeta^\mu}\omega^{\mu\nu}
  \frac{\partial F_2}{\partial \zeta^{\nu}} \, .
\end{equation}

This equation for the quantum Poisson bracket can be put into the
coordinate--independent form
\begin{equation}
  \label{eq:PoissonCommutatorEquality}
  i\hbar \{ F_1 , F_2 \}_{\text{QM}}(\psi) =
  \langle \psi |[\hat{F_1},\hat{F_2}]| \psi \rangle 
\end{equation}
for any two functions $F_1(\psi) = \langle \psi |\hat F_1
|\psi\rangle$ and $F_2(\psi) = \langle \psi |\hat F_2 |\psi\rangle$
given by expectation values of operators.  This expression of the
Poisson bracket has the advantage that it applies to an
infinite--dimensional Hilbert space. 
Moreover, the standard equation for time evolution in quantum
mechanics,
\begin{equation}
  \label{eq:FullQuantumEvolution}
  \dot{F}(\psi) = -{\frac{i}{\hbar}}
  \langle \psi |[\hat{F},\hat{H}]| \psi \rangle \, ,
\end{equation}
again returns the classical--like expression
\begin{equation}
  \dot{F} = \{ F , H \}_{\text{QM}} \, .
\end{equation}

Note that, although this Poisson bracket is of the same form as that
of classical mechanics, it is defined on the whole projective Hilbert
space $\mathbb{PH}$ whose only resemblance to a classical phase space
is that it is a symplectic manifold; in particular, $\mathbb{PH}$ is
generally infinite--dimensional whereas the corresponding classical
phase space is finite.

These results show that there is a homomorphism from the (infinite)
Lie algebra of all Hermitian linear operators on $\mathbb{H}$ to a
corresponding Poisson bracket algebra of functions on a phase space.
However, the resulting dynamics is not classical; as a phase space,
the projective Hilbert space is generally too large.  For the example
of a single particle in one dimension, the Hilbert space is
infinite--dimensional whereas the required classical phase space is
two--dimensional.  In the following section it is shown that, under
certain conditions, the classical dynamics of an algebraic model can
be realized on a submanifold of the projective Hilbert space.  Such
submanifolds are naturally realized as coherent state manifolds.

\section{Classical mechanics as constrained quantum mechanics}
\label{sec:Quantum}

To extract a mechanics from quantum mechanics that approximates ideal
classical mechanics as closely as allowed by the uncertainty
principle, it is appropriate to consider coherent state manifolds of
minimal uncertainty states.  However, while it is important to
consider the limits of precision that are in principle attainable, it
should be recognized that classical mechanics rarely deals
with measurements at the level of quantum uncertainties. 
For example, in a situation in which classical mechanics is considered
useful, it is rare that one could make a measurement of (say) the
magnitude of a system's angular momentum that would be sufficiently
precise to distinguish its discrete quantized values.  Thus, in
situations where a classical description applies, it is not typically
possible to identify a unique unitary irrep to which a given system
might belong.  Indeed, it is generally appropriate to characterize a
system at a finite temperature by mixed states with contributions from
different irreps.

We therefore consider two types of coherent state manifolds.  We first
consider the special case of a submanifold of pure states of the
Hilbert space for an irrep of a dynamical group, and show that it is
possible to constrain the quantal dynamics to such a submanifold,
provided it is symplectic, and thereby obtain a classical dynamics.
The second more general type of manifold we consider is a set of
generalized coherent states comprising arbitrary mixed states
described by density matrices.  We show how classical dynamics is
obtained generally for such a manifold.

\subsection{Coherent state submanifolds}

Let $\mathfrak{g}$ be a SGA for an algebraic system, and $G$ a
corresponding dynamical group with $\mathfrak{g}$ as its Lie algebra.
Let $T$ be an irreducible  unitary representation of $G$ on a Hilbert
space $\mathbb{H}$; together with a Hamiltonian operator $\hat{H}$,
this representation defines a quantal algebraic model.  Any normalized
state vector $|0\rangle\in\mathbb{H}$ defines a classical density
$\rho_m$ for which
\begin{equation}
  \label{eq:ClassicalDensity}
  \rho_m(A) = \langle 0| \hat A|0\rangle , \quad \forall\
  A\in\mathfrak{g}\,.
\end{equation}
Since $G$ acts on $\mathbb{H}$ as a group of canonical
transformations, there exists a moment map $|g\rangle
\equiv T^\dagger (g) |0\rangle \to\rho_{m\cdot g}$, where
\begin{equation}
  \label{eq:SystemCoherentStates}
  \rho_{m\cdot g}(A) =\langle g| \hat A |g\rangle = \rho_m(A(g))\,,
\end{equation}
for all $A\in\mathfrak{g}$, with $A(g)$ defined by Eq.~(\ref{eq:Adg}).
Therefore, because a classical phase space is equivalent to an orbit
$\mathcal{O}_m$ of the dynamical group $G$, it follows that the moment
map takes every $G$--orbit in $\mathbb{PH}$
\begin{equation}
  \mathcal{C} =  \{ |g\rangle = T^\dag(g)|0\rangle ;\ g\in G\}\,,
\end{equation}
to a classical phase space
\begin{equation}
  \mathcal{O}_m = \{ \rho_{m\cdot g} \, ;\, g\in G\} \,.
\end{equation}
Orbits of $G$ in $\mathbb{PH}$ are known as \emph{systems of coherent
  states}~\cite{perelomov}.  Thus, every system of coherent states of
a dynamical group defines a corresponding classical 
phase space  and all its kinematical properties.

Note that the map $\mathcal{C} \to \mathcal{O}_m$ is kinematical; that
is, it relates the two--form on $\mathcal{C}$, given by the
restriction of the Fubini--Study metric of
Eq.~(\ref{eq:DefFubiniStudy}), to the classical symplectic structure
on $\mathcal{O}_m$.  However, if the map $\mathcal{C} \to
\mathcal{O}_m$ is one-to-one, then with the addition of constraints,
it also defines a classical Hamiltonian function $\mathcal{H}$ and a
dynamics on $\mathcal{O}_m$.

Constraints are needed because, in general, the quantal time evolution
of a coherent state does not remain a coherent state.  The time
evolution of a harmonic oscillator coherent state based on the
harmonic oscillator ground state, which evolves classically under
quantal time evolution, is an exception. Other exceptions arise for
systems whose Hamiltonians lie in the SGA. Thus, in general, the
quantal time evolution of a state that is initially a coherent state
does not obey classical equations of motion.  To obtain classical
behaviour, it is necessary to \emph{constrain} the quantal time
evolution to prevent it from leaving the manifold of coherent states.

The theory of constrained Hamiltonian dynamics was first investigated
by Dirac~\cite{dirac64} and generalized in geometric terms by Gotay,
Nester and Hinds~\cite{gotaynesterhinds}.  Rowe, Ryman and
Rosensteel~\cite{rowe80}, for example, applied the theory of
constraints to quantum systems and showed that Dirac's time--dependent
variational principle, Eq.~(\ref{eq:Dirac}), defines a Hamiltonian
dynamics on any symplectic submanifold of $\mathbb{PH}$, i.e., a
submanifold on which the restriction of the symplectic form $\omega$
is nondegenerate~\cite{rowe80,rowe83,Kramer}.  We refer to such a
Hamiltonian dynamics as constrained quantum mechanics.  Thus, a
classical Hamiltonian dynamics is defined by constraining quantum
mechanics to any system of coherent states that is symplectic.

Let $\{ z^\nu\}$ denote a set of coordinates for some neighbourhood in
$\mathcal{C}$.  Then, with the Hamiltonian function $H$ on
$\mathcal{C}$ expressed as a function of these coordinates by
\begin{equation}
  \label{eq:QuantumHamiltonianFunction}
  H(g(z)) = \langle g(z)|\hat H|g(z)\rangle \, ,
\end{equation}
the derivatives of $H$ are given by
\begin{align}
  (\partial_\mu H)(g(z)) &= \langle \partial_\mu g(z) | \hat H
  |g(z)\rangle + \langle g(z)  | \hat H
  |\partial_\mu g(z) \rangle \nonumber \\
  &= - \sigma_{\mu\nu} \dot z^\nu \,,
\end{align}
where $\partial_\mu \equiv \partial/\partial z^\mu$ and $\sigma$ is
the restriction of the Fubini--Study symplectic form $\omega$ on
$\mathbb{PH}$ to $\mathcal{C}$.  

A special case occurs when $\mathcal{C}$ is symplectic.  In this case,
$(\sigma_{\mu\nu})$ can be inverted to give the 
(constrained) equations of motion
\begin{equation}
  \dot{z}^{\mu} ={\sigma}^{\mu\nu} (\partial_\nu H) \,,
  \label{eq:EofM}
\end{equation}
with $(\sigma^{\mu\nu})$ defined such that
\begin{equation}
  \sum_\mu \sigma^{\mu\nu}\sigma_{\mu\lambda} = \delta^\nu_\lambda \,.
\end{equation}
These equations are equivalently obtained from Dirac's variational
principle.  They are expressed in terms of Poisson brackets by
observing that, for any function  $F(z) \in
C^{\infty}(\mathcal{C})$  on the coherent state orbit,
the time evolution is given by
\begin{equation}
  \dot{F} = (\partial_\mu F) \dot z^\mu =  \{ F, H\}_{\mathcal{C}} \,, 
\end{equation}
where
\begin{equation}
  \{ F, H\}_{\mathcal{C}} = (\partial_\mu F)
  {\sigma}^{\mu\nu} (\partial_\nu H) \,.
\end{equation}

If a coherent state submanifold $\mathcal{C}$ of $\mathbb{PH}$ is not
symplectic then, as noted by Rowe \emph{et al.}~\cite{rowe80}, Dirac's
time-dependent variational principle does not define a constrained
quantum dynamics on $\mathcal{C}$; the metric $(\sigma_{\mu\nu})$ does
not have an inverse and Eq.~(\ref{eq:EofM}) is not defined.  However,
even if not symplectic, $\mathcal{C}$ maps to a symplectic coadjoint
orbit under the moment map.  Moreover, it is possible to define a
Hamiltonian function $\bar{H}$ on $\mathcal{O}$ as an average of $H$
over the states of $\mathcal{C}$ that map to a single classical state.
Such an averaging process is defined and rationalized below.  First,
however, we note that if the map $\mathcal{C}\to \mathcal{O}$ is
many--to--one, it cannot be inverted and, hence, no single quantal
state in $\mathcal{C}$ is assigned to a given classical state in
$\mathcal{O}$.  This suggests that one should map a given classical
state in $\mathcal{O}$ to a suitably defined mixture of the quantal
states on $\mathcal{C}$ that cannot be distinguished by a measurement
of classical observables~\cite{rosensteelrowe81}, i.e., by the
expectation values of elements in the SGA.

\subsection{Coherent manifolds of mixed states} 
\label{sect:choice}

Recall that a pure (normalized) state $|0\rangle$ in quantum mechanics
can be represented by a quantal density (often called a density
matrix) $\hat\rho_0 = |0\rangle\langle 0|$ which maps to a classical
density $\rho_0$, defined on the SGA of observables, by $\rho_0 (A) =
{\rm Tr}(\hat \rho_0 \hat A) = \langle 0|\hat A |0\rangle$ for $A\in
\mathfrak{g}$.  Similarly, a mixed state is represented by a density
$\hat{\rho} =\sum_i p_i |i\rangle \langle i|$, where $\{|i\rangle\}$
is a set of pure quantum states and $\{p_i \geq 0\}$ are coefficients
of a probability distribution satisfying $\sum_i p_i = 1$, 
and also gives a classical density $\rho$ defined by
\begin{equation}
  \label{eq:ClassicalMixedDensity}
  \rho(A) = {\rm Tr}(\hat \rho \hat A) = \sum_i p_i \langle i|\hat A
  |i\rangle \, ,
\end{equation}
for $A \in \mathfrak{g}$. 

To consider mixed states that span many irreps, we now allow $T$ to be
a generally reducible unitary representation of $G$. Generalizing
Eqs.~(\ref{eq:ClassicalDensity}) and (\ref{eq:SystemCoherentStates}),
it is seen that an arbitrary density $\hat{\rho}$ for an algebraic
system with dynamical group $G$ defines a corresponding manifold of
mixed coherent states containing $\hat{\rho}$ given by
\begin{equation}
  \label{eq:SystemMixedCoherentStates}
  \mathcal{C} = \{ \hat{\rho}_g = T(g^{-1}) \hat{\rho}\, T(g), g \in G
  \} \, .
\end{equation}
Moreover, there is a moment map, $\mathcal{C} \to\mathcal{O}_m; \hat
\rho_g \mapsto \rho_{m\cdot g}$, in which these coherent states map to
a coadjoint orbit of classical densities with
\begin{equation}
  \label{eq:ClassicalMixedDensity}
  \rho_{m\cdot g} (A)= {\rm Tr}(\hat{\rho}_g\,\hat{A})
  = \sum_i p_i \langle i|T(g)\hat{A}T(g^{-1})|i\rangle\,,
\end{equation}
for all $A$ in the SGA $\mathfrak{g}$ (the Lie algebra of $G$).

A density $\hat\rho_g$ has a natural interpretation as an element of the
dual of the Lie algebra of linear operators on $\mathbb{H}$ with the
standard pairing
\begin{equation}
  \hat\rho_g : \hat X \to {\rm Tr}(\hat{\rho}_g\,\hat{X}) \,.
\end{equation}
The moment map, $\hat\rho_g \to \rho_{m\cdot g}$, is then seen as the
restriction of $\hat \rho_g$ to the operators of the
(generally  reducible) unitary representation
$T$ of the Lie algebra $\mathfrak{g}$ on $\mathbb{H}$, i.e., to
$\{\hat A = T(A); A\in\mathfrak{g}\}$.  It follows that the functions
\begin{equation}
  \mathcal{A}(m\cdot g) = {\rm Tr}(\hat{\rho}_g\,\hat{A})
\end{equation}
and their Poisson brackets
\begin{equation}
  \{\mathcal{A},\mathcal{B}\}(m\cdot g) = -\frac{i}{\hbar}
  \rho_{m\cdot g} ([A,B]) \, ,
\end{equation}
are well-defined on $\mathcal{O}_m$ for all $A,B \in
\mathfrak{g}$.  It also follows that any coherent state
manifold of mixed (or pure) states defines a classical representation
of $\mathfrak{g}$.  This result is non-trivial because, in general,
the coherent state manifold $\mathcal{C}$ is of higher dimension than
the corresponding coadjoint orbit $\mathcal{O}_m$.  As a result, the
moment map does not preserve the commutation relations of arbitrary
linear operators.

Being group orbits, $\mathcal{O}_m$ and $\mathcal{C}$ can be
characterized as coset spaces.  The coadjoint orbit $\mathcal{O}_m
\simeq K_\rho\backslash G$ has stability subgroup
\begin{equation}
  \label{eq:MixedStability}
  K_\rho = \{ h\in G | {\rm Tr}(\hat{\rho}_h\,\hat{A}) =
  {\rm Tr}(\hat{\rho}\,\hat{A}),\; \forall A\in \mathfrak{g}\} \,,
\end{equation}
whereas the stability subgroup for $\mathcal{C} \simeq
K_{\hat\rho}\backslash G$ is
\begin{equation}
  K_{\hat\rho} = \{ h\in G | {\rm Tr}(\hat{\rho}_h\,\hat{X}) =
  {\rm Tr}(\hat{\rho}\,\hat{X})\} \,,
\end{equation}
for any linear operator $\hat X$ that is bounded on $\mathcal{C}$.

To have a dynamics on $\mathcal{O}_m$, we need to know that the
time-derivatives of the expectation values of the observables of
$\mathfrak{g}$ at points of $\mathcal{C}$ defined according to the
standard time-dependent Schr\"odinger equation by
\begin{equation}
  \frac{{\rm d}}{{\rm d}t} {\rm Tr}(\hat{\rho}_g\,\hat{A}) = 
  -\frac{i}{\hbar}\, {\rm Tr}(\hat{\rho}_g\,[\hat{A},\hat H]) \,,
\end{equation}
map to well-defined functions on $\mathcal{O}_m$.  It is seen that
they do provided ${\rm Tr}(\hat{\rho}_{hg}\,[\hat{A},\hat H])$ is
independent of $h$ for all $h\in K_\rho$ and all $A\in\mathfrak{g}$.

\medskip
\noindent {\em Claim:}
If $H(g) ={\rm Tr}(\hat{\rho}_{g}\hat H)$ satisfies $H(hg) = H(g)$,
for all $h\in K_\rho$ and $g\in G$, then $\mathcal{H}(m\cdot g) =
H(g)$ is a well-defined Hamiltonian on $\mathcal{O}_m$ and there is a
well-defined Hamiltonian dynamics on $\mathcal{O}_m$ defined by
\begin{equation}
  \dot{\mathcal{A}}(m\cdot g) = -\frac{i}{\hbar}\, {\rm
Tr}(\hat{\rho}_g\,[\hat{A},\hat H]) \,.
\end{equation}

\noindent {\em Proof:}
Define $\mathcal{H}(m)= {\rm Tr}(\hat{\rho}\hat H)$. Then, if ${\rm
  Tr}(\hat{\rho}_{h}\hat H)={\rm Tr}(\hat{\rho}\hat H)$ for all $h\in
K_\rho$, we can define $\mathcal{H}(m\cdot g)= {\rm
  Tr}(\hat{\rho}_g\hat H)$  which is
well-defined on
$\mathcal{O}_m$.  Now, assuming the conditions of the claim to be
satisfied, we have
\begin{equation}
  \mathcal{H}(m\cdot g e^{-\frac{i}{\hbar} aA}) = 
  {\rm Tr}(\hat{\rho}_{g}e^{-\frac{i}{\hbar} aA}\hat He^{\frac{i}{\hbar}
  aA}) \,
\end{equation}
with $a\in \mathbb{R}$ and $A\in \mathfrak{g}$.
It follows that
\begin{equation}
 -\frac{i}{\hbar}\, {\rm Tr}(\hat{\rho}_g\,[\hat{A},\hat H])
= \frac{\partial\mathcal{H}}{\partial a}(m\cdot g e^{-\frac{i}{\hbar}
aA})\Big|_{a=0}
\end{equation}
is then well-defined on $\mathcal{O}_m$. \hfill  $\emph{QED}$
\medskip

It is seen from the claim that the Hamiltonian on $\mathcal{C}$
automatically maps to a well-defined function on $\mathcal{O}_m$ in
the special situation in which $K_\rho$ and $K_{\hat\rho}$ are the
same and the manifolds $\mathcal{C}\simeq\mathcal{O}_m\simeq
K_\rho\backslash G$ are diffeomorphic.  In general, the Hamiltonian on
$\mathcal{C}$ does not map directly to a well-defined function on
$\mathcal{O}_m$, but it is now clear how to adjust the map so that it
does.  One can define a Hamiltonian
on $\mathcal{O}_m$ as the average
\begin{equation}
  \mathcal{H}(m\cdot g) = \bar H(g) = \frac{1}{v_{K_\rho}}\int_{K_\rho}
  H(hg)\, {\rm d}v(h) \,,
\end{equation}
where ${\rm d}v$ is the left-invariant measure on $K_\rho$ and
$v_{K_\rho}$ is the volume of $K_\rho$ with respect to this measure.
(Note that the averaged energy function $\bar H(g)$ satisfies the
condition of the claim  that  $\bar H(hg) =\bar H(g)$ for all $h\in
K_\rho$.) 

Such an averaging of $H$ is rationalized as follows.  If the map
$\mathcal{C}\to \mathcal{O}_m$ is many--to--one, it cannot be
inverted.  Hence, no  unique  quantal state
(density) in $\mathcal{C}$ is assigned to a given classical state in
$\mathcal{O}_m$.  This multivaluedness suggests that to each classical
state $\rho_{m\cdot g}\in\mathcal{O}_m$ one should assign a new
quantal density $\hat \rho'_g$ corresponding to a mixture with
 a suitable weighting of all the quantal
states in $\mathcal{C}$ that cannot be distinguished by a measurement
of classical observables, i.e., the expectation values of elements in
the SGA. There are many possibilities for choosing suitable mixed
states~\cite{rosensteelrowe81,rosensteel81}. The simplest is to weight
all coherent states of a group orbit in $\mathbb{PH}$ that map to a
single classical state by the invariant measure of the stability
subgroup of the coadjoint orbit.  A map $\mathcal{C}\to\mathcal{C}'$
in which $\hat\rho_g \to\hat\rho'_g$ is then defined by
\begin{equation}
  \label{eq:DensityMatrix}
  \hat\rho'_g = \frac{1}{v_{K_\rho}}\int_{K_\rho}
  \hat\rho_{hg}\, {\rm d}v(h) \,,
\end{equation}
It is clear that the new coherent state manifold $\mathcal{C}'$ is now
diffeomorphic to $\mathcal{O}_m$ and that the moment map
$\mathcal{C}'\to \mathcal{O}_m$ defines a classical Hamiltonian
dynamics.

A (potential) problem arises with the above construction if it should
happen that the stability subgroup ${K_{\rho}}$ is non-compact.  The
volume $v_{K_\rho}$ is then infinite and the above expressions for
$\mathcal{H}$ and $\hat \rho'_g$ are not defined. The problem is
resolved if the volume $v_{K_{\rho} / K_{\hat{\rho}}}$ of the factor
space $K_{\rho} / K_{\hat{\rho}}$ is finite.
Eq.~(\ref{eq:DensityMatrix}) can then be replaced by
\begin{equation}
  \hat{\rho}'_g = \frac{1}{v_{K_{\rho} / K_{\hat{\rho}}} }
  \int_{K_{\rho} / K_{\hat{\rho}}} \hat{\rho}_{hg} \, {\rm d}v(h) \,,
\end{equation}
where ${\rm d}v$ is now the $K_{\rho}$--invariant measure on this
factor space.

\section{An Example:  The Asymmetric Top}
\label{sec:Rotor}

The rigid rotor provides insightful examples of the procedures
developed in this paper.  Despite their apparent simplicity,
rotational models are considerably richer than a traditional canonical
problem with three degrees of freedom.  In particular, their phase
spaces have non--trivial geometries and admit the possibility of
intrinsic degrees of freedom.  It will be shown that the rigid rotor
has a natural spectrum generating algebra and its constrained quantum
mechanics yield classical equations of motion.

\subsection{A spectrum generating algebra for the asymmetric top}

A rigid rotor is characterized by three \emph{intrinsic} moments of
inertia $(\overline{\Im}_1, \overline{\Im}_2, \overline{\Im}_3)$ which
are its moments of inertia in the intrinsic (principal axes) frame of
reference.  We assume the three moments of inertia to be all
different; the rotor is then known as an \emph{asymmetric
  top}~\cite{townes}.  Because the intrinsic moments of inertia of the
rotor are fixed (the rigidity condition), the observables of the rotor
depend only its orientation and angular momentum.  The orientation of
an asymmetric top is characterized by an inertia tensor $\Im$,
whose moments $\Im_{ij}$ (in a Cartesian basis) are the elements of a
real symmetric $3 \times 3$ matrix.  Given the values of the inertia
tensor, the corresponding orientation of the rotor is then defined by
the rotation matrix $\Omega \in SO(3)$ that brings the inertia tensor
to diagonal form,
\begin{equation}
  \label{eq:DefinitionInertiaTensor}
  \Im_{ij} =  [\tilde\Omega \overline{\Im} \Omega]_{ij} =
  \sum_k \Omega_{ki} \overline{\Im}_k \Omega_{kj}  \, , 
\end{equation}
where $\overline{\Im}$ is the diagonal inertia tensor in the intrinsic
frame with diagonal elements $(\overline{\Im}_1, \overline{\Im}_2,
\overline{\Im}_3)$ and $\tilde\Omega$ is the transpose of $\Omega \in
SO(3)$.  Because the inertia tensor is a function only of orientation,
the components $\Im_{ij}$ commute,
\begin{equation}
  \label{eq:InertiaTensorCommutes}
  [\Im_{ij},\Im_{kl}] = 0 \, ,
\end{equation}
and span an algebra isomorphic to $\mathbb{R}^6$.

The angular momentum $\mathbf{L}$ has Cartesian components $\{ L_i;
i=1,2,3 \}$ which span an $so(3)$ Lie algebra,
\begin{equation}
  \label{eq:so(3)LieAlgebra}
  [L_i,L_j] = i \hbar \, L_k, \quad i,j,k\ {\rm cyclic.}
\end{equation}
The inertia tensor defined by (\ref{eq:DefinitionInertiaTensor}) is a
rank--2 Cartesian tensor.  Thus, it obeys the commutation relations
\begin{equation}
  \label{eq:ImIsRank2SphericalTensor}
  [\Im_{ij},L_k] = i \hbar \sum_l (\varepsilon_{lik} \Im_{lj} +
  \varepsilon_{ljk} \Im_{li} ) \, .
\end{equation}
Together, the moments of inertia and the angular momenta span a SGA
for the rotor that is isomorphic to the semidirect sum algebra
$[\mathbb{R}^6]so(3)$ with $\mathbb{R}^6$ as its ideal.  This algebra
is known as the \emph{rotor model algebra} (RMA)~\cite{rowe96}.

\subsection{The quantum asymmetric top} 

In quantum mechanics, the elements of the RMA are interpreted as the
Hermitian linear operators of an irreducible unitary represention. In
view of Eq.~(\ref{eq:DefinitionInertiaTensor}), it is clear that the
the moments of inertia $\{\Im_{ij}\}$ can be represented by the linear
operators $\{\hat\Im_{ij}\}$ on $\mathcal{L}^2(SO(3))$ defined by
\begin{equation}
  [\hat\Im_{ij}\Psi](\Omega) =  [\tilde\Omega \overline{\Im}
  \Omega]_{ij} \Psi(\Omega) \,.
\end{equation}
Similarly, the angular momentum operators are  represented
in the usual way as infinitesimal generators of
rotations, where a finite rotation of a function in
$\mathcal{L}^2(SO(3))$ is defined by
\begin{equation}
  [R(\Omega)\Psi](\Omega') = \Psi(\Omega'\Omega) \,.
\end{equation}
Howeover, it is known from the theory of induced
representations~\cite{mackey} that the Hilbert space
$\mathcal{L}^2(SO(3))$ is reducible.  This reducibility can be
inferred from the fact that the configuration space for the asymmetric
top is not $SO(3)$ but the factor space $D_2\backslash SO(3)$, where
$D_2 \subset SO(3)$ is the subgroup of all elements of $SO(3)$ that
leave the inertia tensor in the intrinsic frame invariant;
\begin{equation}
  D_2 =\{ \omega \in SO(3)\, |\, \tilde\omega \overline{\Im}
  \omega =  \overline{\Im} \}\, .
\end{equation}
The subgroup $D_2$ is the group generated by rotations through $\pi$
about the principal axes of the inertia tensor.  It is a discrete
group known in crystallography as the \emph{dihedral group}.  It has
four one--dimensional unitary irreps, $\{ \chi^{(i)}; i = 1, \ldots ,
4\}$, and one two--dimensional spinor unitary irrep $\chi^{(5)}$.
Thus, as known from the theory of induced
representations~\cite{mackey} (cf.\ also \cite{rowe96}), a unitary
irrep of the RMA is defined on the subspace of functions in
$\mathcal{L}^2 (SO(3))$ that satisfy the condition
\begin{equation}
  \Psi(\omega\Omega) = \chi^{(i)}(\omega)
  \Psi(\Omega), \quad \forall \, \omega\in D_2 \, ,
\end{equation}
for one of the irreducible one--dimensional unitary representations of
$D_2$.  A spinor unitary irrep of the RMA is similarly defined on a
subspace of half--odd integer functions in $\mathcal{L}^2(SU(2))$.
 Let $\mathbb{PH}^{(i)}$ denote the projective Hilbert
space that carries the irrep of the RMA induced from $\chi^{(i)}$.

The quantum (kinetic energy) Hamiltonian is defined to be
\begin{equation}
  \label{eq:QuantumHamiltonianOperator}
  \hat{H} = \tfrac{1}{2} \sum_{mn} \hat{L}_m \hat{\Im}^{-1}_{mn}
  \hat{L}_n \, ,
\end{equation}
where $\{ \hat \Im_{mn}\}$ and $\{\hat L_m\}$ are, respectively, the
moments of inertia and angular momentum operators in the above
representations.

Further details of this and other quantum rigid rotors can
be found in~\cite{rowe96}. 

\subsection{Coherent state manifolds for the rigid rotor}

The elements of the RMA in the above representations are the
infinitesimal generators of a \emph{rotor model group} (RMG); the
angular momentum operators generate rotations and the moments of
inertia generate angular momentum boosts.  A coherent state manifold
for the rigid rotor is generated by acting on a state $\,|0\rangle \in
\mathbb{PH}^{(i)}$ with the representation of
the RMG given by exponentiation of the RMA as
\begin{multline}
  \label{eq:RMGCSmanifold}
  \mathcal{C} = \Bigl\{\, |Q,\Omega\rangle =
  R(\tilde\Omega) \exp \big[\frac{i}{\hbar}
  \sum_{ij} Q_{ij} \hat\Im_{ij} \big] |0\rangle \, ;
  \\ Q \in [\mathbb{R}^6]\, ,\, \Omega\in SO(3) \Bigr\} \,.
\end{multline}
A desirable choice of $|0\rangle\in\mathbb{PH}^{(i)}$ is a state for
which the expectation values of the rotor observables are equal to the
values they would have for an ideal classical state with well--defined
orientation and zero angular momentum.  The state $|0\rangle$ can be
chosen such that
\begin{equation}
  \langle {0}| \hat{L}_k | {0} \rangle = 0 \, , \quad k = 1,2,3,
\end{equation}
and all polynomials in the Lie algebra of odd degree in the $so(3)$
subalgebra have vanishing expectation value.  This choice is possible
because there is an involution on the algebra $[\mathbb{R}^6]so(3)$,
having the physical interpretation of time reversal, in which
\begin{equation}
  {\hat{\Im}}_{ij} \to {\hat{\Im}}_{ij} \, , \quad
  \hat L_k \to -\hat L_k \, .
\end{equation}
Thus, by choosing $|0\rangle$ to be invariant under time reversal, all
expectation values of polynomials in the algebra that are odd under
time reversal are zero.  It is also possible to choose $|0\rangle$
such that the expectation of the inertia tensor $\Theta_{ij} = \langle
0 | {\hat{\Im}}_{ij}|0\rangle$ is diagonal, i.e., $\Theta_{ij} =
\delta_{ij}\Theta_i$, and the values $\Theta_i$ are
arbitrarily close to the intrinsic moments of inertia,
i.e., $\Theta_i \approx\overline{\Im}_i$.  By suppressing the quantum
mechanical uncertainty in orientation, these states possess a
correspondingly large uncertainty in the angular momentum.

The expression (\ref{eq:RMGCSmanifold}) for a coherent state, with a
time-reversal invariant $|0\rangle$, then gives
\begin{align}
  \Im_{ij}(\Omega) &= \langle Q,\Omega | \hat{\Im}_{ij} | Q,\Omega
  \rangle \nonumber \\
  \label{eq:ClassicalInertiaTensor}
  &= (\tilde\Omega \Theta \Omega)_{ij} \approx (\tilde\Omega
  \overline{\Im} \Omega)_{ij} \, , \\ 
  \mathcal{L}_k(Q,\Omega) &= \langle Q,\Omega | \hat{L}_k | Q,\Omega
  \rangle \nonumber \\
  &= \sum_l \bar{\mathcal{L}}_l(Q) \Omega_{lk} \,,   
  \label{eq:ClassicalAngularMomentum}
\end{align}
where 
\begin{equation}
  \bar{\mathcal{L}}_l(Q)  = 
  \frac{i}{\hbar} \langle 0| [\hat{L}_l , \hat{Q}]|0\rangle = \sum_{ijl}
  Q_{ij} (\bar{\Im}_i -\bar{\Im}_j) \varepsilon_{ijl}  \label{eq:Ibar}
\end{equation}
and $\hat Q = \sum_{ij} Q_{ij} \hat{\Im}_{ij}$.

\subsection{Constraining the rotor mechanics}

The Hilbert space of the rotor does not contain
normalizable eigenstates of any moment of inertia: an eigenstate can
only be approached as a delta function limit in which the
uncertainties in the orientation become negligible.  Thus,
 the isotropy subgroup  $S^{(i)}$
of the coherent state manifold at  $|0\rangle
\in \mathbb{PH}^{(i)}$ cannot contain any element of
$\mathbb{R}^6$ (other than the identity).  For a true
(one--dimensional) representation, it is found that 
$S^{(i)} = D_2$,  and for a two--dimensional spinor
representation $S^{(5)}\to\bar S$ is a
subgroup of $\bar D_2$, the double covering of $D_2$.  Thus, in both
cases, the coherent state manifold is 9--dimensional and, being of odd
dimension, it cannot be symplectic.

The isotropy subgroup $H \subset [\mathbb{R}^6]SO(3)$ of the
corresponding coadjoint orbit can be determined by expressing the
observables as functions of a set of 9 coordinates for the coherent
state manifold and seeing which ones are redundant on the coadjoint
orbit.  A suitable set of coordinates for $\mathcal{C}$ about
$|0\rangle$ is given by the 6 coefficients $Q_{ij}$ in the expansion
of $Q = \sum_{ij} Q_{ij} \Im_{ij}$ and a set of three coefficients
$\xi^k$ of the angular momenta in the expansion $\Omega = \exp (
-\frac{i}{\hbar} \sum _k \xi^k L_k )$.  Thus, for $(Q,\Omega)$ close
to $(0,I_3)$, where $I_3$ is the identity in $SO(3)$, the moments of
inertia $\Im_{ij}$ are functions of all the $\xi^k$.  However, the
angular momenta, $\mathcal{L}_k(Q,\Omega)$, are independent of the
diagonal coordinates $\{ Q_{ii}\}$.  It follows that the isotropy
subgroup of the coadjoint orbit is a subgroup $[\mathbb{R}^3]S \subset
[\mathbb{R}^6]SO(3)$ whose infinitesimal generators are the diagonal
moments $\{ \Im_{ii}\}$ of the inertia tensor.  The coadjoint orbit
$\mathcal{O}$ corresponding to the coherent state manifold, then, has
the geometry of $[\mathbb{R}^3]S\backslash [\mathbb{R}^6]SO(3)$; it is
6--dimensional, symplectic, and 
diffeomorphic to the cotangent bundle $T^*( S\backslash SO(3))$. The
classical functions $\Im_{ij}(\Omega)$ and $\mathcal{L}_k(Q,\Omega),$
given by Eq.~(\ref{eq:ClassicalInertiaTensor}) and
(\ref{eq:ClassicalAngularMomentum}), satisfy a Poisson bracket algebra
isomorphic to the RMA:
\begin{align}
  \{ \Im_{ij} , \Im_{kl} \} &= 0 \, , \\
  \{ \mathcal{L}_i , \mathcal{L}_j \} &= \sum_k \varepsilon_{ijk}
  \mathcal{L}_k \, , \\
  \{ \Im_{ij} , \mathcal{L}_k \} &= \sum_l (\varepsilon_{lik}
  \Im_{lj} + \varepsilon_{ljk} \Im_{li}) \, .
\end{align}

This model is an example of a quantum system whose coherent state
manifolds are not symplectic and therefore not diffeomorphic to
classical phase spaces.  Nevertheless, they project naturally to
coadjoint orbits which are.  One is then led to enquire if the
constrained dynamics is well--defined on a coadjoint orbit
$\mathcal{O}$.  Of particular concern is whether or not the
expectation of the Hamiltonian $\hat{H}$ for coherent states depends
on the gauge degrees of freedom associated with the $[\mathbb{R}^3]S$
subgroup.  The rotor Hamiltonian is expressed in terms of the RMA as
\begin{equation}
  \label{eq:QuantumHamiltonianOperator2}
  \hat{H} = \tfrac{1}{2} \sum_{mn} \hat{L}_m \hat{\Im}^{-1}_{mn}
  \hat{L}_n \, ,
\end{equation}
and the corresponding energy function on $\mathcal{C}$ is given by
\begin{equation}
  \label{eq:RotorHamiltonianFunction}
  H(Q,\Omega) = \langle Q,\Omega | \hat{H} |Q,\Omega \rangle \, .
\end{equation}

With the expression (\ref{eq:RMGCSmanifold}) for a coherent state, the
value of the energy function is given by
\begin{equation}
  H(Q,\Omega) = E_0 - \frac{1}{2\hbar^2}
  \langle 0 |[[ \hat H,\hat Q],\hat Q] | 0 \rangle \, ,
\end{equation}
where $E_0= \langle 0 | \hat H | 0 \rangle$ and we have used the fact
that because of Eq.~(\ref{eq:ImIsRank2SphericalTensor}), the
commutator $[\hat H , \hat Q]$ is odd in the angular momentum
operators and so has vanishing expectation value in the
time--reversal invariant  state $|0\rangle$.
Also, because $[[\hat L_m,\hat Q],\hat Q] = 0$, the energy is given by
\begin{equation}
 H(Q,\Omega) = E_0 - \frac{1}{2\hbar^2}  
  \sum_{mn} \langle 0|[\hat L_m,\hat Q]\, \hat{\Im}^{-1}_{mn}\, [\hat
  L_n,\hat Q]|0\rangle \, .
\end{equation}

Now, if the state $|0 \rangle$ were an eigenstate of the moments of
inertia, i.e., $\hat{\Im}_{ij} |0\rangle = \delta_{ij}
\overline{\Im}_i |0\rangle$, then, by
Eq.~(\ref{eq:ImIsRank2SphericalTensor}), it would also be an
eigenstate of $[\hat L_n,\hat Q] $ with eigenvalue $ -i\hbar
\bar{\mathcal{L}}_n$, where $\bar{\mathcal{L}}_n\equiv
\bar{\mathcal{L}}_n(Q)$ is defined by Eq.~(\ref{eq:Ibar}). We would
then have
\begin{equation}
  - \frac{1}{2\hbar^2}  \sum_{mn} \langle 0| [\hat L_m,\hat
  Q]\, \hat{\Im}^{-1}_{mn}\, [\hat L_n,\hat Q]|0\rangle =
  \mathcal{H}(\Im, \mathcal{L}) \,,
\end{equation}
where $\mathcal{H}(\Im, \mathcal{L})$ is the ideal classical Hamiltonian
\begin{equation}
  \mathcal{H}(\Im, \mathcal{L}) = \tfrac{1}{2} \sum_{m} \bar{\mathcal{L}}_m
  \bar{\Im}^{-1}_{m}\bar{\mathcal{L}}_m
  =\tfrac{1}{2} \sum_{mn} \mathcal{L}_m {\Im}^{-1}_{mn}\mathcal{L}_n \,. 
\end{equation} 

Unfortunately, eigenstates of the moments of inertia are not
normalizable and not in the Hilbert space. Nevertheless, it is
possible to define sequences of normalizable states, which approach
eigenstates in the limit, and for which all $\langle
0|\hat\Im_{ij}|0\rangle$ become arbitrarily close to
$\delta_{ij}\overline{\Im}_i$; i.e.,
\begin{equation}
  \lim\, \langle{0}|{\hat{\Im}}_{ij}|{0}\rangle = \delta_{ij}
  \overline{\Im}_i\,.
\end{equation}
A state $|0\rangle$ for which this limit is approached has a
relatively sharp orientation and a correspondingly large uncertainty
in its angular momentum state, in accord with the constraints of the
uncertainly principle.

It follows that the expectation of $\hat H$, in the limit of coherent
states with precisely--defined orientations, gives the energy function
\begin{equation}
  \lim H(Q,\Omega) = E_0 + \mathcal{H}(\Im, \mathcal{L}) \, .
\end{equation}
Thus $H \to E_0 + \mathcal{H}$ becomes well--defined on the coadjoint
orbit (without need for averaging in the limit) and leads to the
standard classical equations of motion for the rotor (i.e., Euler's
equations),
\begin{align}
  \label{eq:ClassicalRotorEquations}
  \dot{\Im}_{ij} &= \{ \Im_{ij} , \mathcal{H} \} \, , \\
  \dot{\mathcal{L}}_k &= \{ \mathcal{L}_k , \mathcal{H} \} \, .
\end{align}

It should be understood that while the above limiting procedure gives
the idealised classical mechanics of a rotor, it can be approached but
never quite achieved in practice.  However by following the procedures
outlined in this paper and by taking coherent states that involve a
small but finite uncertainty in both the angular momentum and the
inertia tensor as the classical embedding, one obtains a classical
mechanics that is consistent, to within some (classical) uncertainty,
with the idealised classical model.

\section{Conclusions}
\label{sec:Conclusions}

It is shown in this paper that constraining the quantal dynamics of an
algebraic model to an appropriately embedded coherent state manifold
leads to classical equations of motion for a Hamiltonian that might be
observed with a physical distribution of ideal classical states. Such
a classical dynamics is consistent with but not, generally, identical
to the corresponding, unachievable, ideal classical dynamics which
takes no account of the uncertainty principle.  There are many ways to
embed a classical phase space in a quantum model.  The embeddings that
most closely approximate the idealized classical mechanics are
provided by coherent state orbits of minimal uncertainty states for
which the measurable observables are defined as precisely as possible
to within the constraints of the uncertainty principle.  Then, since
the map from the coherent states to classical densities may still not
be one-to-one, we propose an averaging over the (classically
unobservable) gauge degrees of freedom.  More generally, we provide
embeddings in which a classical state is represented by a mixed
quantum state spanning many irreps.

The circumstances under which a quantal system should behave
classically, i.e., when constrained quantum mechanics should be an
adequate replacement for the unconstrained mechanics, is not
investigated.  However, the results of this paper give new insights
into how classical behaviour may appear in a quantum system.
Ballentine~\cite{Bal94,Bal98} has numerically investigated regimes
where (unconstained) quantum dynamics gives compatible results to
distributions of classical states.  Such compatible results can
presumably be anticipated only in regimes where a classical
description is expected to apply.  In conclusion, it is emphasized
that our purpose in exploring the route from quantum mechanics to
classical mechanics is to understand better the conditions that must
be satisfied by an acceptable theory of quantization, i.e., a theory
which prescribes a route in the opposite direction
\cite{scalar,vector}.

\begin{acknowledgments}
  We thank C Bahri and J Repka for useful discussions.  SDB
  acknowledges the support of a Macquarie University Research
  Fellowship.  This paper was supported in part by a Macquarie
  University Research Grant and by NSERC of Canada.
\end{acknowledgments}


\begin{thebibliography}{99}

\bibitem{diracbook}
  P. A. M. Dirac, \textit{The Principles of Quantum Mechanics,} (Oxford
  University Press, Oxford, 1958).

\bibitem{scalar} 
  S. D. Bartlett, D. J. Rowe, and J. Repka, J. Phys. A:
  Math. Gen. \textbf{35}, 5599 (2002).

\bibitem{vector} 
  S. D. Bartlett, D. J. Rowe, and J. Repka, J. Phys. A:
  Math. Gen. \textbf{35}, 5625 (2002).

\bibitem{mackey} 
  G. W. Mackey, Ann. of Math. \textbf{55}, 101 (1952); 
  \textit{Induced Representation of Groups and
  Quantum Mechanics,} (Benjamin, New York, 1968); 
  \textit{Unitary Group Representations in Physics, Probability and
  Number Theory,} (Benjamin, Reading, MA, 1978).

\bibitem{kostant} 
  B. Kostant, \textit{On Certain Unitary Representations which Arise
  from a Quantization Theory}, in \textit{Group Representations in
  Mathematics and Physics, Lecture Notes in Physics, Vol. 6}
  (Springer, Berlin, 1970); \textit{Quantization and
  Unitary Representations}, in \textit{Letures in Modern Analysis and
  Applications III, Lecture Notes in Mathematics, Vol.\ 170}
  (Springer, Berlin, 1970).

\bibitem{souriau} J.--M. Souriau, Comm. Math. Phys.  
  \textbf{1}, 374 (1966);  \textit{Structure des syst\`emes
  dynamiques,} (Dunod, Paris, 1970).

\bibitem{woodhouse} N. M. J. Woodhouse, \textit{Geometric
  Quantization} (Oxford University Press, Oxford, 1991).

\bibitem{joseph}
  A. Joseph, Commun. Math. Phys. \textbf{17}, 210 (1970);
  M. J. Gotay, in \textit{Quantization, coherent states, and complex
  structures, (Bialowieza 1994),} (Plenum, New York, 1995);
  M. J. Gotay, H. B. Grundling, and C. A. Hurst, 
  Trans. Amer. Math. Soc. \textbf{348}, 1579 (1996);
  M. J. Gotay and H. B. Grundling, Rep. Math. Phys. \textbf{40}, 107 (1997).

\bibitem{rowebook}
  D. J. Rowe, \textit{Nuclear Collective Motion,} (Methuen, London, 1970).

\bibitem{TDHF} 
  D. J. Rowe and R. Basserman, Can. J. Phys. \textbf{54}, 1941 (1976).

\bibitem{rowe80}
  D. J. Rowe, A. Ryman, and G. Rosensteel, \pra \textbf{22}, 2362 (1980).
      
\bibitem{rowe83}
  D. J. Rowe, M. Vassanji, and G. Rosensteel, \pra \textbf{28},
  1951, (1983). 
                                
\bibitem{Ehr27} 
  P. Ehrenfest, Z. Phys. \textbf{45}, 455 (1927). 

\bibitem{Bal94}
  L. E. Ballentine, Y. Yang and J. P. Zibin, \pra \textbf{50}, 2854
  (1994).

\bibitem{Teg01} 
  M. Tegmark and J. A. Wheeler, ``100 Years of Quantum
  Mysteries,'' Scientific American, Feb. (2001), pp. 68-75.

\bibitem{marsden94}
  J. E. Marsden and T. S. Ratiu, \textit{Introduction to Mechanics and
  Symmetry,} (Springer/Verlag, New York, 1994).

\bibitem{ihrig}
  E. Ihrig and G. Rosensteel, Int. J. Theoret. Phys. \textbf{32}, 843
  (1993). 

\bibitem{SGA}
  A. O. Barut, A. Bohm, and Y. Ne'eman, eds. \textit{Dynamical Groups
  and Spectrum Generating Algebras,} (World Scientific, Singapore, 1986).

\bibitem{gotay}
  M. J. Gotay, J. Math. Phys. \textbf{40}, 2107 (1999).

\bibitem{perelomov}
  A. Perelomov, Commun. Math. Phys. \textbf{26}, 222 (1972);
  \textit{Generalized Coherent States and their Applications,} (Springer,
  Berlin, 1986).

\bibitem{dirac64}
  P. A. M. Dirac, \textit{Lectures on Quantum Mechanics,} (Belfer Graduate
  School of Science Monograph Series 2, 1964).

\bibitem{gotaynesterhinds}
  M. J. Gotay, J. M. Nester, and G. Hinds, J. Math. Phys.
  \textbf{19}, 2388 (1978).

\bibitem{Kramer} 
  P. Kramer and M. Saraceno, \textit{Geometry of the
  time-dependent variational principle in quantum mechanics,}
  (Springer-Verlag, Berlin, 1981).

\bibitem{rosensteelrowe81}
  G. Rosensteel and D. J. Rowe, \pra \textbf{24}, 673 (1981).
  
\bibitem{rosensteel81}
  G. Rosensteel, \pra \textbf{23}, 2794 (1981). 

\bibitem{townes} 
  C. H. Townes and A. L. Schawlow, \textit{Microwave Spectroscopy,}
  (Dover, New York, 1975).

\bibitem{rowe96}
  D. J. Rowe, Prog. Part. Nucl. Phys. \textbf{37}, 265 (1996).

\bibitem{Bal98} 
  L. E. Ballentine and S. M. McRae, \pra \textbf{58}, 1799 (1998).

\end{thebibliography}
\end{document}